\documentclass{amsart} %{amsart}%{article}%{elsart}
\usepackage{graphicx}
\usepackage{amssymb, amsmath}
\usepackage{amsfonts}
\usepackage{amssymb}
\usepackage{float}

\newtheorem{theorem}{Theorem}

\newtheorem{definition}[theorem]{Definition}

\newcommand{\C}{\mathbb C}

\newcommand{\R}{\mathbb R}

\def\la{\label}
\def\bt{\begin{thm}}
\def\et{\end{thm}}

\def\bl{\begin{lem}}
\def\el{\end{lem}}

\def\bd{\begin{defi}}
\def\ed{\end{defi}}

\def\bc{\begin{cor}}
\def\ec{\end{cor}}

\def\bp{\begin{proof}}
\def\ep{\end{proof}}

\def\br{\begin{rem}}
\def\er{\end{rem}}

\newtheorem{thm}{Theorem}[section]
\newtheorem{lem}{Lemma}[section]
\newtheorem{defi}{Definition}[section]

\newtheorem{rem}{Remark}[section]
\newtheorem{cor}{Corollary}[section]

\numberwithin{equation}{section}
\numberwithin{theorem}{section}
\numberwithin{example}{section}
\numberwithin{figure}{section}

\begin{document}
\title{Dynamic Phase Transitions for Ferromagnetic Systems}
\author[Ma]{Tian Ma}
\address[TM]{Department of Mathematics, Sichuan University,
Chengdu, P. R. China}

\author[Wang]{Shouhong Wang}
\address[SW]{Department of Mathematics,
Indiana University, Bloomington, IN 47405}
\email{showang@indiana.edu}

\thanks{The work was supported in part by the
Office of Naval Research and by the National Science Foundation.}
\thanks{{\tt http://www.indiana.edu/~fluid}}


\begin{abstract}
This article presents a phenomenological dynamic phase transition theory for ferromagnetism, leading to a  precise description of the dynamic transitions, and to a physical predication on the spontaneous magnetization. The analysis also suggests asymmetry of fluctuations in  both the  ferromagnetism and the PVT systems.
\end{abstract}
\maketitle
%\tableofcontents

\section{Introduction}
\label{sc1}
Classically, phase transitions are classified by  the Ehrenfest classification scheme,  based on 
 the lowest derivative of the free energy that is discontinuous at the transition. 
For ferromagnetic  systems, it has been observed that the magnetization, which is the first derivative of the free energy with the applied magnetic field strength, increases continuously from zero as the temperature is lowered below the Curie temperature, and the  magnetic susceptibility, the second derivative of the free energy with the field, changes discontinuously. 
Hence, the ferromagnetic phase transition in materials such as iron is regarded as a  second order phase transition. However, a theoretical understanding of the transition is still lacking. The main objective of this article is to provide theoretical approach to dynamic phase transitions for ferromagnetic systems.

%Two models based on the classical  and the revised Ginzburg-Landau free energies. are examined. 
For the classical GL free energy, although both the steady state and time-dependent models provide some results in agreement with experiments, there are obvious discrepancies on both susceptibility and spontaneous magnetization. 
Hence a revised GL free energy is proposed and analyzed, leading to 
a precise description of the dynamic transitions, and to a physical predication on the spontaneous magnetization. 
 
The analysis  is based on the recently developed dynamic transition theory by the authors, together with a new dynamic  classification scheme, which classifies  phase transitions into three categories: Type-I, Type-II and Type-III, corresponding mathematically  to continuous,  jump, and  mixed transitions, respectively; see Section 2 and the Appendix as well as two recent books by the authors \cite{b-book, chinese-book} for details. 
 
We remark also that  the analysis leads naturally to a physical conjecture on  asymmetry of fluctuations, which appears in both  the  ferromagnetic system studied in this article  and in PVT systems  studied in \cite{mw-pvt}.

This article is organized as follows. In Section 2, we review the dynamic classification scheme and the new time-dependent Ginzburg-Landau model for equilibrium phase transitions. Section 3 deals with the dynamic transition based on the classical 
Ginzburg-Landau energy, and Section 4 addresses the dynamic transition theory using a revised Ginzburg-Landau energy. Physical conclusions are given in Section 5, and dynamic transition theory is recapitulated in the Appendix.

\section{General Principles of Phase Transition Dynamics}
In this section,  we recapitulate the new dynamic phase transition classification scheme 
to classify phase transitions into three categories: Type-I, Type-II and Type-III, corresponding mathematically  to continuous,  jump, and  mixed transitions, respectively. 

Then  we recall a new time-dependent Ginzburg-Landau theory for modeling  equilibrium phase transitions in statistical physics,  derived based on the le Ch\^atelier principle and some mathematical insights on pseudo-gradient systems. 

Both the classification scheme and the Ginzburg-Landau theory was developed recently by the authors, and we refer interested readers to \cite{b-book, chinese-book, mw-pvt} for details.

\subsection{Dynamic classification scheme}
In sciences, nonlinear dissipative systems are generally governed
by differential equations, which can be expressed in the following
abstract form 
\begin{equation}
\left.
\begin{aligned} 
&\frac{du}{dt}=L_{\lambda}u+G(u,\lambda ),\\
&u(0)=\varphi , 
\end{aligned} \right.\label{7.1}
\end{equation}
where $u:[0,\infty )\rightarrow X$ is the unknown function,
$\lambda\in \R^N$ $(N\geq 1)$ is  the control parameter, $X$  and $X_1$ are two Banach
spaces with $X_1\subset X$ being a  dense and compact inclusion,
$L_{\lambda}=-A+B_{\lambda}$ and $G(\cdot ,\lambda
):X_1\rightarrow X$ are $C^r(r\geq 0)$ mappings depending
continuously on $\lambda$,  $L_{\lambda}:X_1\rightarrow X$ is a
sectorial operator, and
\begin{equation}
\left.
\begin{aligned} &A:X_1\rightarrow X  && \text{a\ linear\
homeomorphism},\\
&B_{\lambda}:X_1\rightarrow X&& \text{a\ linear\ compact\
operator}.
\end{aligned}
\right.\label{7.2}
\end{equation}

In following, we introduce some basic and universal concepts in
nonlinear sciences.  

First, a  state of the system (\ref{7.1}) at $\lambda$ is usually referred to as a compact invariant set $\Sigma_{\lambda}$. In many applications, 
$\Sigma_{\lambda}$ is a singular point or a periodic orbit. A state $\Sigma_{\lambda}$
of (\ref{7.1}) is stable if $\Sigma_{\lambda}$ is an attractor,
otherwise $\Sigma_{\lambda}$ is called unstable.

Second,  we say that the system (\ref{7.1}) has a
phase transition from a state $\Sigma_{\lambda}$ at $\lambda
=\lambda_0$ if $\Sigma_{\lambda}$ is stable on $\lambda <\lambda_0$
(or on $\lambda >\lambda_0$) and is unstable on $\lambda
>\lambda_0$ (or on $\lambda <\lambda_0$).  The critical parameter $\lambda_0$ is
called a critical point.
 In other words, the
phase transition corresponds to an exchange of stable states.

The concept of phase transition originates from the statistical
physics and thermodynamics. In physics and chemistry, "phase"
means the homogeneous part in a heterogeneous system. However,
here the so called phase means the stable state in the systems of
nonlinear sciences including physics, chemistry, biology, ecology,
economics, fluid dynamics and geophysical fluid dynamics, etc.
Hence, here the content of phase transition has been endowed with
more general significance. In fact, the phase transition dynamics
introduced  here can be applied to a wide variety of topics involving
the universal critical phenomena of state changes in nature in a
unified mathematical viewpoint and manner.

Third, if the system (\ref{7.1}) possesses the gradient-type structure, then the phase transitions are called equilibrium phase transition; otherwise they are called the non-equilibrium phase transitions.

Fourth, classically, there are several ways to classify phase transitions. The one most used  is the Ehrenfest classification scheme, which groups phase transitions based on the degree of non-analyticity involved.  First order phase transitions are  also called discontinuous, and  higher order phase transitions $(n>1)$ are called continuous.

Here we introduce the following notion of  dynamic classification scheme: 

\begin{definition}%[Dynamic Classification of Phase Transition]
Let $\lambda_0\in \R^N$ be a critical point of (\ref{7.1}), and (\ref{7.1}) undergo a transition from state $\Sigma^1_{\lambda}$ to $\Sigma^2_{\lambda}$. There are three types of phase transitions for  (\ref{7.1}) at $\lambda =\lambda_0$, depending on their  dynamic properties: continuous, jump, and mixed as given in Theorem~\ref{t5.1}, which are called Type-I, Type-II and Type-III respectively.
\end{definition}

The main characteristics  of Type-II phase transitions is that there is a gap between $\Sigma^1_{\lambda}$ and $\Sigma^2_{\lambda}$ at the critical point $\lambda_0$. In thermodynamics,  the metastable states correspond in general to the super-heated or super-cooled states, which have been found in many physical phenomena. In particular,
Type-II phase transitions are always accompanied with the latent heat to occur.

In a Type-I phase transition, two states $\Sigma^1_{\lambda}$ and 
$\Sigma^2_{\lambda}$ meet  at $\lambda_0$, i.e., the system undergoes 
a continuous transition from $\Sigma^1_{\lambda}$ to
$\Sigma^2_{\lambda}$.

In a Type-III phase transition, there are at least two different stable states $\Sigma^{\lambda}_2$ and $\Sigma^{\lambda}_3$ at $\lambda_0$, and system undergoes a continuous transition to $\Sigma^{\lambda}_2$ or a jump transition  to $\Sigma^{\lambda}_3$, depending on the  fluctuations.

It is clear that a Type-II phase transition of gradient-type
systems must be discontinuous or the zero order because there is a
gap between $\Sigma^1_{\lambda}$ and $\Sigma^j_{\lambda}$  $(2\leq j\leq
K)$. For a Type-I phase transition, the energy  is continuous, and consequently, it is an $n$-th order transition in the Ehrenfest sense  for some $n\geq 2$.
A Type-III phase transition is indefinite, for the transition from
$\Sigma^1_{\lambda}$ to $\Sigma^2_{\lambda}$ it may be continuous, i.e.,
$dF^-/d\lambda =dF^+_2/d\lambda$, and for the transitions from
$\Sigma^1_{\lambda}$ to $\Sigma^3_{\lambda}$ it may be discontinuous:
$dF^-/d\lambda\neq dF^+_3/d\lambda$ at $\lambda =\lambda_0$.

\subsection{Time-dependent Ginzburg-Landau models for equilibrium phase transitions}
%\subsection{Thermodynamic potentials}
\label{s7.2.2}
In this subsection, we introduce the time-dependent Ginzburg-Landau model for equilibrium phase transitions. 

We start with thermodynamic potentials and the Ginzburg-Landau free energy. As we know, four thermodynamic potentials-- internal energy, the enthalpy, the Helmholtz free energy and the Gibbs free energy--are useful in the chemical thermodynamics of reactions and non-cyclic processes. 

Consider a thermal system, its order parameter $u$ changes in
$\Omega\subset \R^n$ $(1\leq n\leq 3)$. In this situation,  the free energy of this system is of the form
%\begin{widetext}
\begin{equation}
{\mathcal{H}}(u,\lambda
)= {\mathcal{H}}_0 
 +\int_{\Omega}\Big[\frac{1}{2}\sum^m_{i=1}\mu_i|\nabla
u_i|^2  +g(u,\nabla u,\lambda )\Big]dx\label{7.28}
\end{equation}
%\end{widetext}
where $N\geq 3$ is an integer,
$u=(u_1,\cdots,u_m),\mu_i=\mu_i(\lambda )>0$, and $g(u,\nabla
u,\lambda )$ is a $C^r(r\geq 2)$ function of $(u,\nabla u)$ with
the Taylor expansion \begin{equation} g(u,\nabla u,\lambda
)=\sum\alpha_{ijk}u_iD_ju_k+\sum^N_{|I|=1}\alpha_Iu^I+o(|u|^N)-fX,\label{7.29}
\end{equation}
where $I=(i_1,\cdots,i_m),i_k\geq 0$ are integer,
$|I|=\sum^m_{k=1}i_k$, the coefficients $\alpha_{ijk}$ and
$\alpha_I$ continuously depend on $\lambda$, which are determined
by the concrete physical problem, $u^I=u^{i_1}_1\cdots u^{im}_m$
and $fX$ the generalized work.

Thus, the study of thermal equilibrium phase transition for the
static situation is referred to the steady state bifurcation of
the system of elliptic equations
$$\left.
\begin{aligned}
&\frac{\delta}{\delta u}{\mathcal{H}}(u,\lambda )=0,\\
&\frac{\partial u}{\partial n}|_{\partial\Omega}=0,\ \ \ \
(\text{or}\ u|_{\partial\Omega}=0),
\end{aligned}
\right.$$ where $\delta /\delta u$  is the variational derivative.

A thermal system is controled by some parameter $\lambda$. When
$\lambda$ is for from the critical point $\lambda_0$ the system
lies on a stable equilibrium state $\Sigma_1$, and when $\lambda$
reaches or exceeds $\lambda_0$ the state $\Sigma_1$ becomes unstable, and 
meanwhile the system will undergo a transition  from $\Sigma_1$ to another stable
state $\Sigma_2$. The basic principle is that there often exists  fluctuations in the system leading  to a deviation from the equilibrium states, and  the phase transition process is a
dynamical behavior, which should be described by a time-dependent
equation.

To derive a general time-dependent model, first we recall that  the classical  
le Ch\^atelier  principle amounts to saying that 
for a stable  equilibrium state of a system $\Sigma$, when the system deviates from
$\Sigma$ by a small perturbation or fluctuation, there will be a
resuming force to retore this system to return to the stable state
$\Sigma$.
Second, we know that  a stable equilibrium state of a thermal system must
be the minimal value point of the thermodynamic potential. 

By the mathematical characterization of gradient systems and the le Ch\^atelier principle, for a system with
thermodynamic potential ${\mathcal{H}}(u,\lambda )$, the governing
equations are essentially determined by the functional
${\mathcal{H}}(u,\lambda )$.
When the order parameters $(u_1,\cdots,u_m)$ are nonconserved
variables, i.e., the integers
$$\int_{\Omega}u_i(x,t)dx=a_i(t)\neq\text{constant}.$$
then the time-dependent equations are given by
\begin{equation}
\left.
\begin{aligned} 
&\frac{\partial u_i}{\partial
t}=-\beta_i\frac{\delta}{\delta u_i}{\mathcal{H}}(u,\lambda
)+\Phi_i(u,\nabla u,\lambda ),\\
&\frac{\partial u}{\partial n}|_{\partial\Omega}=0\ \ \ \
(\text{or}\ u|_{\partial\Omega}=0),\\
&u(x,0)=\varphi (x),
\end{aligned}
\right.\label{7.30}
\end{equation}
for any $1 \le i \le m$, 
where $\delta /\delta u_i$ are the variational derivative,
$\beta_i>0$ and $\Phi_i$ satisfy
\begin{equation}
\int_{\Omega}\sum_i\Phi_i\frac{\delta}{\delta
u_i}{\mathcal{H}}(u,\lambda )dx=0.\label{7.31}
\end{equation}
The condition (\ref{7.31})  is  required by
the Le Ch\^atelier principle. In the concrete problem, the terms
$\Phi_i$ can be determined by physical laws and (\ref{7.31}).

When the order parameters are the number density and the system
has no material exchange with the external, then $u_j$  $(1\leq j\leq
m)$ are conserved, i.e.,
\begin{equation}
\int_{\Omega}u_j(x,t)dx=\text{constant}.\label{7.32}
\end{equation}
This conservation law requires a continuous equation
\begin{equation}
\frac{\partial u_j}{\partial t}=-\nabla\cdot J_j(u,\lambda
),\label{7.33}
\end{equation}
where $J_j(u,\lambda )$ is the flux of component $u_j$. In
addition, $J_j$ satisfy
\begin{equation}
J_j=-k_j\nabla (\mu_j-\sum_{i\neq j}\mu_i),\label{7.34}
\end{equation}
where $\mu_l$ is the chemical potential of component $u_l$, 
\begin{equation}
\mu_j-\sum_{i\neq j}\mu_i=\frac{\delta}{\delta
u_j}{\mathcal{H}}(u,\lambda )-\phi_j(u,\nabla u,\lambda
), \label{7.35}
\end{equation}
and  $\phi_j(u,\lambda )$ is a function depending on the other
components $u_i$ $(i\neq j)$. When $m=1$, i.e., the system consists of
two components $A$ and $B$, this term $\phi_j=0$. Thus, from
(\ref{7.33})-(\ref{7.35}) we obtain the dynamical equations as
follows
\begin{equation}
\left.
\begin{aligned} &\frac{\partial u_j}{\partial
t}=\beta_j\Delta\left[\frac{\delta}{\delta
u_j}{\mathcal{H}}(u,\lambda )-\phi_j(u,\nabla u,\lambda )\right],\\
&\frac{\partial u}{\partial n}|_{\partial\Omega}=0,\ \ \ \
\frac{\partial\Delta u}{\partial n}|_{\partial\Omega}=0,\\
&u(x,0)=\varphi (x),
\end{aligned}
\right.\label{7.36}
\end{equation}
for $1 \le j \le m$, 
where $\beta_j>0$ are constants, $\phi_j$ satisfy
\begin{equation}
\int_{\Omega}\sum_j\Delta\phi_j\cdot\frac{\delta}{\delta
u_j}{\mathcal{H}}(u,\lambda )dx=0.\label{7.37}
\end{equation}

If the order parameters $(u_1,\cdots,u_k)$ are coupled to the
conserved variables $(u_{k+1},\cdots,u_m)$, then the dynamical
equations are
\begin{equation}
\left.
\begin{aligned} 
&\frac{\partial u_i}{\partial t}
   =-\beta_i\frac{\delta}{\delta u_i}{\mathcal{H}}(u,\lambda)+\Phi_i(u,\nabla u,\lambda ),\\
& \frac{\partial u_j}{\partial t}
  =\beta_j\Delta\left[\frac{\delta}{\delta u_j}{\mathcal{H}}(u,\lambda )
    -\phi_j(u,\nabla u,\lambda )\right],\\
&\frac{\partial u_i}{\partial n}|_{\partial\Omega}=0\ \ \ \
(\text{or}\ u_i|_{\partial\Omega}=0),\\
&\frac{\partial u_j}{\partial n}|_{\partial\Omega}=0,\ \ \ \
\frac{\partial\Delta u_j}{\partial n}|_{\partial\Omega}=0,\\
&u(x,0)=\varphi (x).
\end{aligned}
\right.\label{7.38}
\end{equation}
for $1 \le i \le k$  and $k+1 \le j \le m$.

The model (\ref{7.38}) gives a general form of the governing
equations to thermodynamic phase transitions.  Hence, the dynamics of
equilibrium phase transition in statistic physics is based on the new Ginzburg-Landau formulation  (\ref{7.38}).

Physically, the initial value condition $u(0)=\varphi$ in
(\ref{7.38}) stands for the fluctuation of system or perturbation
from the external. Hence, $\varphi$ is generally small. However,
we can not exclude the possibility of a bigger noise $\varphi$.

From conditions (\ref{7.31}) and (\ref{7.37}) it follows that a
steady state solution $u_0$ of (\ref{7.38}) satisfies
\begin{equation}
\left.
\begin{aligned} 
&  \Phi_i(u_0,\nabla u_0,\lambda )=0 && \forall 1\leq
i\leq k,\\
& \Delta\phi_j(u_0,\nabla u_0,\lambda )=0 && \forall k+1\leq j\leq m.
\end{aligned}
\right.\label{7.39}
\end{equation}
Hence a stable equilibrium state must reach the
minimal value of thermodynamic potential. In fact, $u_0$ fulfills
\begin{equation}
\left.
\begin{aligned} 
&
\beta_i\frac{\delta}{\delta u_i}{\mathcal{H}}(u_0,\lambda )
   -\Phi_i(u_0,\nabla u_0,\lambda)=0,\\
&\beta_j\Delta\frac{\delta}{\delta u_j}{\mathcal{H}}(u_0,\lambda)
   -\Delta\phi_j(u_0,\nabla u_0,\lambda )=0,\\
&\frac{\partial u_i}{\partial n}|_{\partial\Omega}=0\ \ \ \
(\text{or}\ u_i|_{\partial\Omega}=0),\\
\\
&\frac{\partial u_j}{\partial n}|_{\partial\Omega}=0,\ \ \ \
\frac{\partial\Delta u_j}{\partial n}|_{\partial\Omega}=0,
\end{aligned}
\right.\label{7.40}
\end{equation}
for $1 \le i \le k$ and $k+1 \le j \le m$.
Multiplying $\Phi_i(u_0,\nabla u_0,\lambda )$ and
$\phi_j(u_0,\nabla u_0,\lambda )$ on the first and the second
equations of (\ref{7.40}) respectively, and integrating them, then
we infer from (\ref{7.31}) and (\ref{7.37}) that
\begin{eqnarray*}
&&\int_{\Omega}\sum_i\Phi^2_i(u_0,\nabla u_0,\lambda )dx=0,\\
&&\int_{\Omega}\sum_j|\nabla\phi_j(u_0,\nabla u_0,\lambda
)|^2dx=0,
\end{eqnarray*}
which imply that (\ref{7.39}) holds  true.

%\section{Ferromagnetism}

\section{Classical Theory  of Ferromagnetism}

A ferromagnetic material consists of lattices containing particles
with a magnetic moment. When no external field is present  and the temperature is above some critical value, called the Curie temperature, the
magnetic moments are oriented at random and there is no net
magnetization. However, as the temperature is lowered, magnetic
interaction energy between lattice sites becomes more important
than the random thermal energy. Below the Curie temperature, the
magnetic moments become ordered in the space and a spontanous
magnetization appears. The phase transition from a paramagnetic to
a ferromagnetic system takes place at the Curie Temperature.

The phase diagrams for magnetic systems are given in Figures
\ref{f8.50}-\ref{f8.52}. In Figure \ref{f8.50}, below the Curie temperature, the
magnetization occurs spontaneously, and the zero magnetic field
$H=0$ separates the two possible orientations of magnetization.
Figure \ref{f8.51} provides a sketch of the isotherms of magnetic system,
and Figure \ref{f8.52} gives the magnetization as a function of
temperature; see also  Reichl \cite{reichl} and Onuki \cite{onuki} for details.
\begin{figure}[hbt]
  \centering
  \includegraphics[width=.6\textwidth]{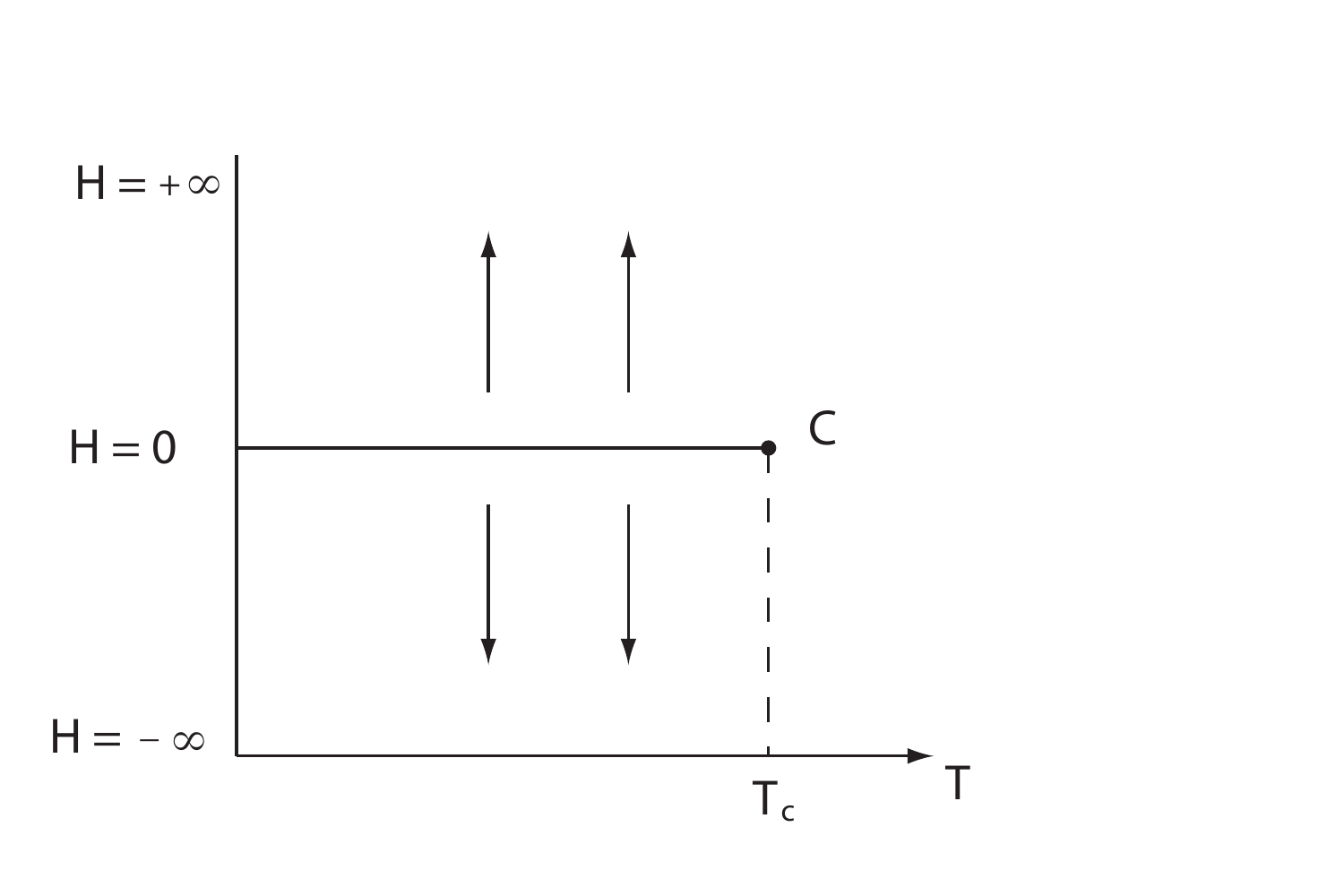}
  \caption{Below the Curie point the magnetization occurs
spontaneously; the curve $H=0$ separates the two possible
orientations of magnetization.}\la{f8.50}
 \end{figure}

\begin{figure}[hbt]
  \centering
  \includegraphics[width=.5\textwidth]{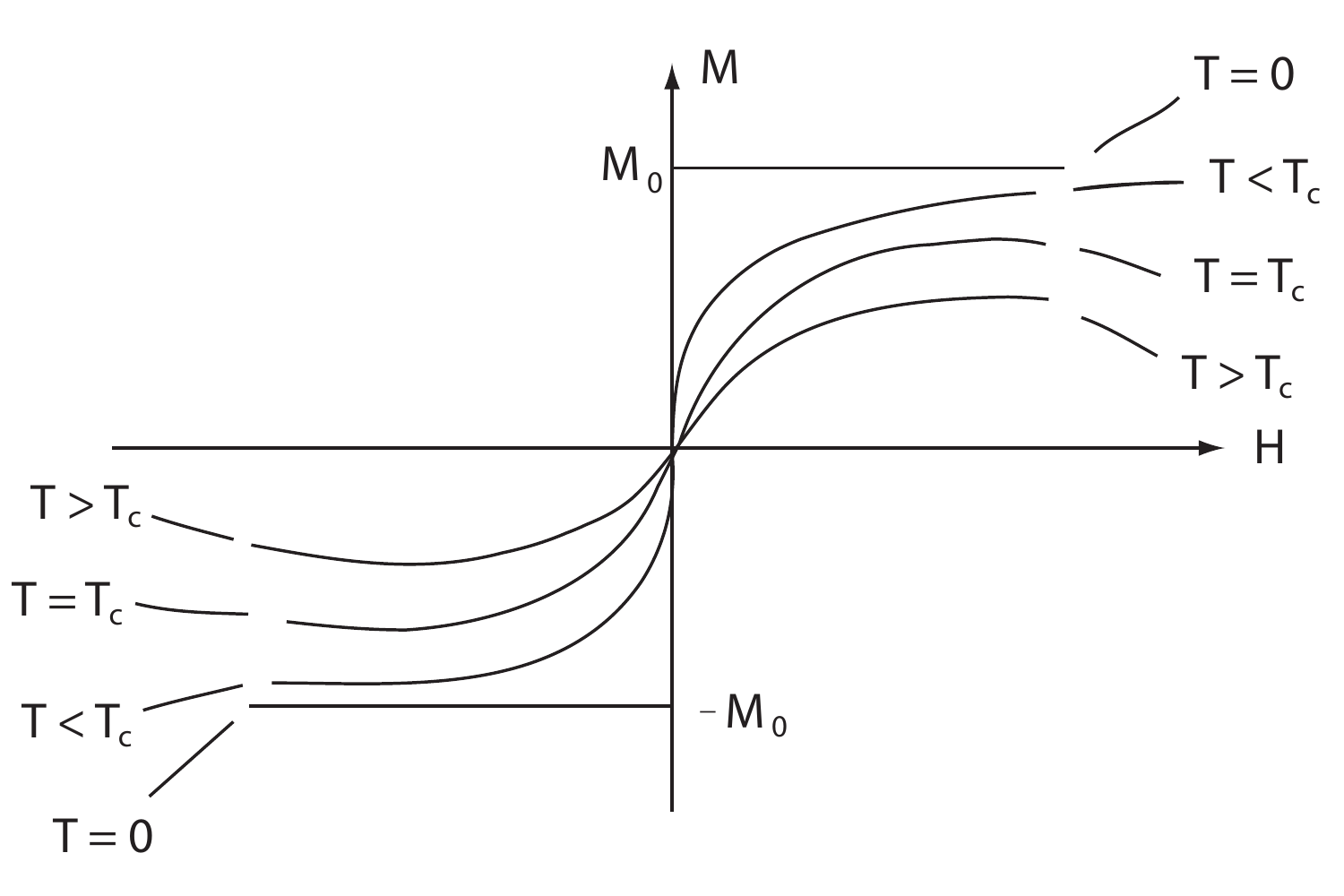}
  \caption{A sketch of the isotherms for a magnetic
system.}\la{f8.51}
 \end{figure}

\begin{figure}[hbt]
  \centering
  \includegraphics[height=.3\textwidth]{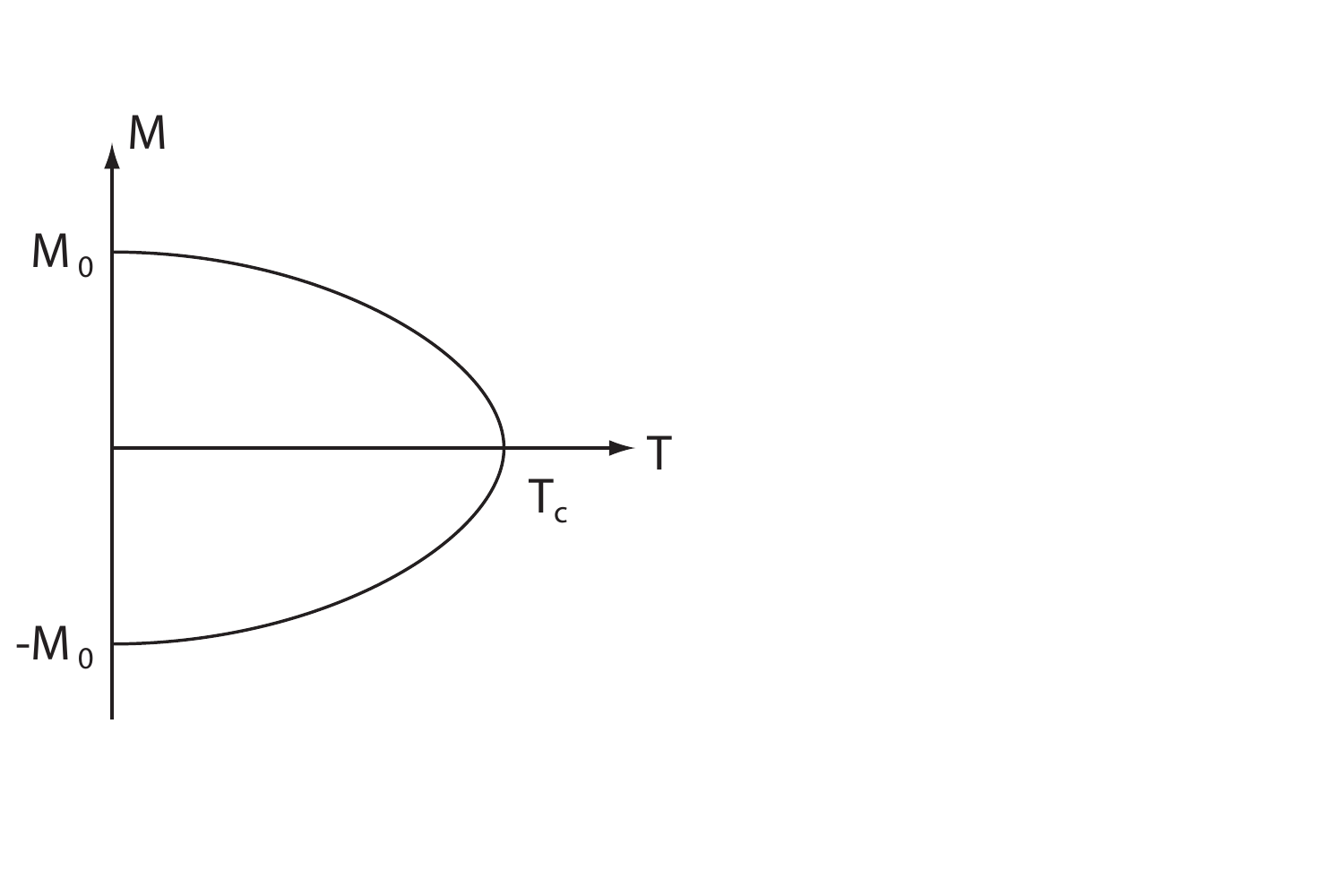}
  \caption{A sketch of the spontaneous magnetization of a
magnetic system.}\la{f8.52}
 \end{figure}

Based on the classical Ginzburg-Landau theory, for an isotropic
system,  the Helmholtz free energy can be expressed as
$$A(M,T)=A_0(T)+\frac{1}{2}\alpha_2(T)|M|^2+\frac{1}{4}\alpha_4(T)|M|^4+\cdots
,$$ 
where $A_0(T)$ is a magnetization-independent contribution to
the free energy, $|M|^2=M\cdot M$, and $M=(M_1,M_2,M_3)$ is the
magnetization of the system. When an external  field $H$ is present, the
Gibbs free energy is  given by
\begin{eqnarray*}
G(M,H,T)&=&A(M,T)-H\cdot M\\
&=&A_0(T)-H\cdot
M+\frac{1}{2}\alpha_2(T,H)|M|^2+\frac{1}{4}\alpha_4(T,H)|M|^4
+\cdots .
\end{eqnarray*}
For small $H$, $\alpha_2$ and $\alpha_4$ can be considered to be
independent of $H$, and near the Curie point $T_c$ we have
$$\alpha_2(T)=\alpha_0(T-T_c),\ \ \ \ \alpha_4(T)>0.$$

Usually, $G(M,H,T)$ is called the Ginzburg-Landau free energy.
To omit the higher order terms than $|M|^4$, it is known that the
equilibrium state $M$ of the ferromagnetic system satisfies
\begin{equation}
\frac{\delta}{\delta
M}G=\alpha_4|M|^2M+\alpha_2M-H=0.\label{8.284}
\end{equation}
Thus above the Curie point we obtain from (\ref{8.284}) that
\begin{equation}
\left.
\begin{aligned} &M\simeq\frac{1}{\alpha_2}H,\\
&\chi =\frac{\partial M}{\partial
H}=\frac{1}{\alpha_2(T)}=\frac{1}{\alpha_0(T-T_c)},
\end{aligned}
\right.\label{8.285}
\end{equation}
where $\chi$ is the isothermal susceptibility, which is a scalar 
because the system is isotropic. Below the critical point, for $H=0$, 
the magnetization $M$ obeys
\begin{equation}
|M|=\sqrt{\frac{\alpha_0(T_c-T)}{\alpha_4}},\qquad \frac{\partial |M|}{\partial
T}=-\frac{1}{2}\sqrt{\frac{\alpha_0}{\alpha_4(T_c-T)}}.
\label{8.286}
\end{equation}
The heat capacity at $T=T_c$ is
\begin{align}
C(T<T_c)-C(T>T_c) =&-T \frac{\partial^2G}{\partial T^2}\Big|_{T=T_c}\label{8.287}\\
=&-T_c\frac{\partial^2}{\partial T^2}\left(\frac{1}{2}\alpha_2|M|^2+\frac{1}{4}\alpha_4|M|^4\right)\Big|_{T=T_c}\nonumber\\
=&  \frac{\alpha^2_0T_c}{2\alpha_4}.\nonumber
\end{align}

We infer then from (\ref{8.285})-(\ref{8.287}) the following  classical
conclusions for an isotropic magnetic system:

\begin{itemize}

\item[(1)] When as external magnetic field is present, a nonzero
magnetization exists above the Curie point $T_c$, which has the
same direction as the applied field $H$.

\item[(2)] Near the critical point $T_c$ the susceptibility
$\chi$ tends to infinite with the rate $(T-T_c)^{-1}$, i.e., a very
small applied field at $T=T_c$ can yield a large effect on the
magnetization.

\item[(3)] In the absence of an external field (i.e., $H=0)$,
below the critical point a spontaneous magnetization $M$ appears,
which depends continuously on $T$ and tends to zero with the rate
$(T-T_c)^{{1}/{2}}$; namely the transition is  of the second order.

\item[(4)] The heat capacity at $T=T_c$ has a jump with the gap
$\Delta C=\frac{\alpha^2_0T_c}{2\alpha_4}$, and the jump has the
shape of a $\lambda$, as shown in Figure \ref{f8.53}.
\end{itemize}

\begin{figure}[hbt]
  \centering
  \includegraphics[width=.4\textwidth]{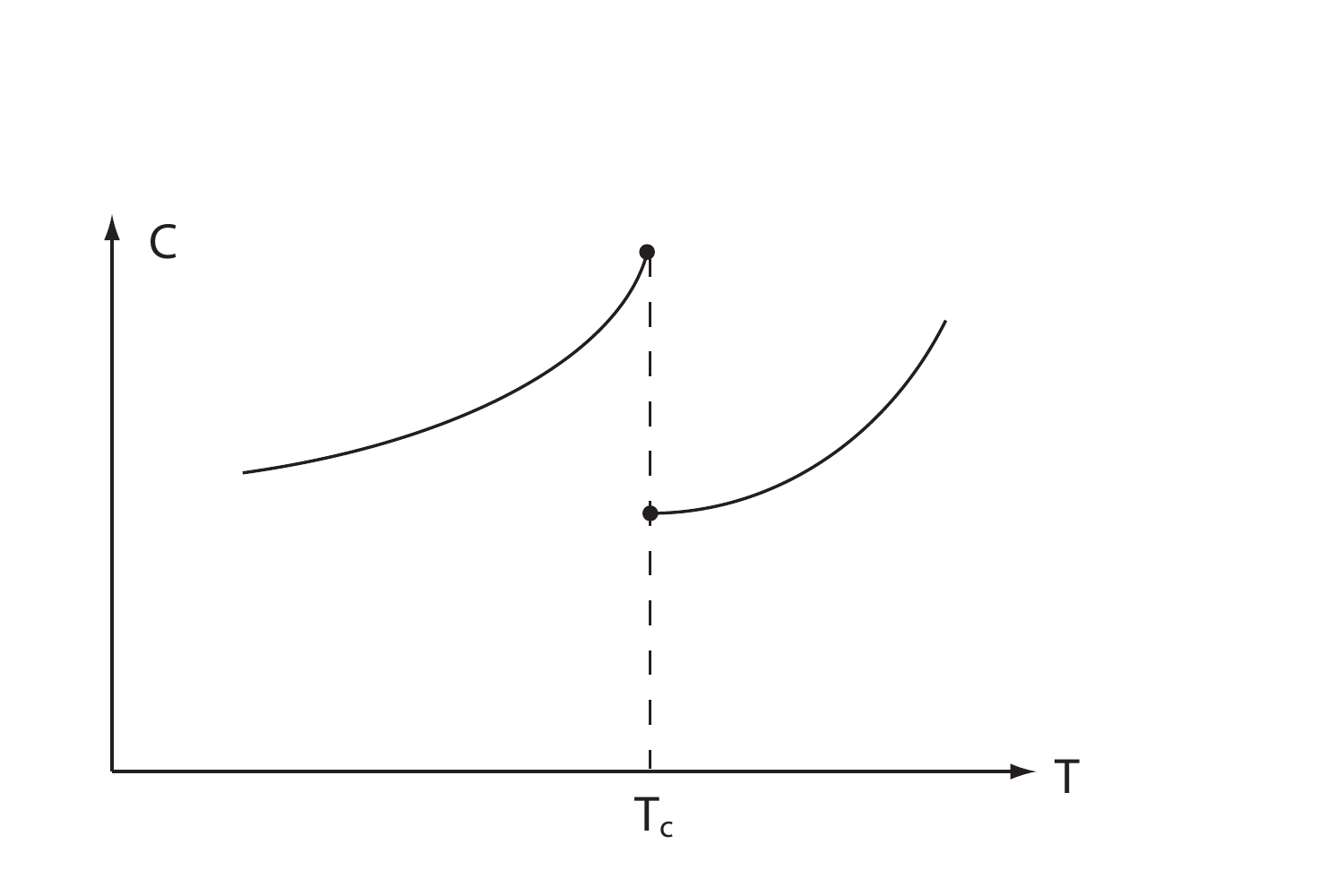}
  \caption{}\la{f8.53}
 \end{figure}

Qualitatively,  part of the above conclusions are in agreement  with
experimental results. 

However these conclusions lead to wrong susceptibility and spontaneous 
magnetization, whose experimental rates are given by 
 $\chi\propto
(T-T_c)^{-r}$ with $r=1.3$, and  by
$M\propto (T-T_c)^{\beta}$ with $\beta ={1}/{3}$.

Free energy $G$ must be a function of the magnetization $M$. Hence
the errors are originated from the fact that the expression of $G$ in the
Ginzburg-Landau theory is an approximation. It is difficult to
derive a precise formula because $G$ is not analytic on $|M|$,
even the differentibility of $G$ on $|M|$ is very low.

If we study the dynamical properties of ferromagnetic systems by
using the classical Ginzburg-Landau free energy, we shall see a
more serious error when an external field
is present.

To see this, the dynamic equation of classical theory is given by
\begin{equation}
\frac{dM}{dt}=-\alpha_2M-\alpha_4|M|^2M+H.\label{8.288}
\end{equation}

For simplicity, we take $H=(h,0,0)$ with $h>0$, it is equivalent
that we take the $x_1$-axis in the direction of $H$. Then the
equation
$$\alpha_4|M|^2M+\alpha_2M-H=0$$
has a steady state solution $M_0=(m_0,0,0)$ for $T\geq 0$, which is
the magnetization induced by $H$. Make the transformation
$$M=M^{\prime}+M_0.$$
Then, the equation (\ref{8.288}) is rewritten as (drop the primes)
\begin{equation}
\left.
\begin{aligned}
&\frac{dM_1}{dt}=-(\alpha_2+3\alpha_4m^2_0)M_1-2\alpha_4m_0M^2_1-\alpha_4m_0|M|^2-\alpha_4|M|^2M_1,\\
&\frac{dM_2}{dt}=-(\alpha_2+\alpha_4m^2_0)M_2-2\alpha_4m_0M_1M_2-\alpha_4|M|^2M_2,\\
&\frac{dM_3}{dt}=-(\alpha_2+\alpha_4m^2_0)M_3-2\alpha_4m_0M_1M_3-\alpha_4|M|^2M_3.
\end{aligned}
\right.\label{8.289}
\end{equation}
Comparing the two critical parameter curves
\begin{align*}
\alpha_2+3\alpha_4m^2_0=2&  \Rightarrow
T_1=T_c-3\alpha_4m^2_0/\alpha_0,\\
\alpha_2+\alpha_4m^2_0=0& \Rightarrow
T_2=T_c-\alpha_4m^2_0/\alpha_0,\end{align*}
 we find that $T_2>T_1$. By
Theorem \ref{t5.1},  (\ref{8.289}) has the first
transition at $T=T_2$, where a new magnetization
$M=(M_1,M_2,M_3)$,   with $M_2\not=0$  and $M_3\not=0$, appears. 
This is unrealistic because any magnetization $M$ of this system
must have the same direction as $H=(h,0,0)$; see Figure \ref{f8.50}.

In fact, when a magnetic field $H$ is applied on an isotropic
system the direction of $H$ is a favorable one for magnetization.
However, in (\ref{8.288}) this point is not manifested. Therefore,
to investigate the phase transition dynamics of ferromagnetic
systems we need to revise the $GL$ free energy.

\section{Dynamic Transitions in Ferromagnetism}

\subsection{Revised Ginzburg-Landau free energy}
Let the ferromagnetic system be isotropic. When a magnetic field
$H$ is present, we introduce a second order symmetric tensor
$$
A(T,H)=(a_{ij}(T,H)),\ \ \ \ a_{ij}=a_{ji},\ \ \ \ 1\leq i,\ j\leq 3,
$$ 
such that $A(T,0)=0$, and  $A(T,H)$ has eigenvalues
$$
\lambda_1=\lambda_1(T,H),\ \ \ \ \lambda_2=\lambda_3=0,\ \ \ \
\text{with}\ \lambda_1(T,H)>0\ \text{as}\ H\neq 0,
$$ 
and $H$ is the
eigenvector of $A$ corresponding to $\lambda_1$:
\begin{equation}
AH=\lambda_1H\ \ \ \ (\lambda_1>0\ \text{as}\ H\neq 0,\ \ \ \
\lambda_1=0\ \text{as}\ H=0).\label{8.290}
\end{equation}
It is clear that if  we take the coordinate system $(x_1,x_2,x_3)$
with $x_1$-axis in the  $H$-direction, then $H=(H_1,0,0)$
and
\begin{equation}
A=\left(\begin{matrix} \lambda_1&0&0\\
0&0&0\\
0&0&0
\end{matrix}
\right).\label{8.291}
\end{equation}

Physically, condition (\ref{8.290}) means that $H$ is a favorable
direction of magnetization if we add a  term 
$$-MAM^T=-\sum
a_{ij}M_iM_j$$ 
in the free energy. We also need to consider the
nonlinear effect acted by $H$. To this end we introduce the term
$-|M|^2M\cdot H$ in the free energy.

Thus when the applied field $H$ may vary in $\Omega\subset
\R^n$  $(n=2,3)$, then the $GL$ free energy is in the form
\begin{align}
G(M,T,H)=G_0&+\frac{1}{2}\int_{\Omega}\Big[\mu |\nabla
M|^2+\alpha_2|M|^2+\frac{\alpha_4}{2}|M|^4   \label{8.292}
\\
&  -\sum a_{ij}M_iM_j -f(T,H,M)M\cdot H \Big]dx,\nonumber
\end{align}
where $G_0=G_0(T)$  is independent of $M$ and $H$, and $f$ is a
scalar function of $T,H$ and $M$,  defined by
\begin{equation}
f(T,H,M)=2(1+\beta |M|^2),\qquad  \beta =\beta
(T,H)>0.\label{8.293}
\end{equation}
For $\alpha_2$ and $\alpha_4$ we  assume that
\begin{equation}
\alpha_2=\alpha_0(H)(T-T_0(H)),\ \ \ \ T_0(0)=T_c,\ \ \ \
\alpha_0>0,\ \ \ \ \alpha_4>0.\label{8.294}
\end{equation}

By the standard model (\ref{7.30}), we derive from (\ref{8.292}) and
(\ref{8.293})  the following dynamical equations:
\begin{align}
\frac{\partial M_i}{\partial t}=& \mu\Delta
M_i-\alpha_2M_i+\sum^3_{j=1}a_{ij}M_j-\alpha_4|M|^2M_i \label{8.295}\\
& +\beta
|M|^2H_i 
+2\beta (M\cdot H)M_i+H_i && \text{for } 1\leq i\leq 3.\nonumber
\end{align}
The boundary condition is given by
\begin{equation}
\left. \frac{\partial M}{\partial
n}\right|_{\partial\Omega}=0.\label{8.296}
\end{equation}
Obviously, if  $H=0$, (\ref{8.295}) coincide with the classical
equations. For simplicity, hereafter we always take
\begin{equation}
H=(h,0,0) \qquad    (h>0\ \text{is a constant}).\label{8.297}
\end{equation}

When $H$ is constant, in the study of phase transitions of
magnetic systems, (\ref{8.295}) can be replaced by a system of
ordinary differential equations as follows:
\begin{equation}
\left.
\begin{aligned}
&\frac{dM_1}{dt}=(\lambda_1-\alpha_2)M_1+\beta h|M|^2+2\beta
hM^2_1-\alpha_4|M|^2M_1+h,\\
&\frac{dM_2}{dt}=-\alpha_2M_2+2\beta hM_1M_2-\alpha_4|M|^2M_2,\\
&\frac{dM_3}{dt}=-\alpha_2M_3+2\beta hM_1M_3-\alpha_4|M|^2M_3.
\end{aligned}
\right.\label{8.298}
\end{equation}

Equations (\ref{8.298}) have a steady state solution induced by $H$:
$$
M^*=(m_0,0,0),\ \ \ \ \text{with}\ \lim_{h \to 0} m_0=  0.
$$ 
We see in Figure \ref{f8.51}
that the magnetization $M^*$ has an upper  bound. Namely there is an
$M_0$ such that
\begin{equation}
\left.
\begin{aligned} 
& |M^*|<|M_0| &&  \qquad  \forall H\in \R^3,\ \ \
\ T\geq 0, \\
& M^*\rightarrow M_0&& \qquad \text{if }  h\rightarrow\infty.
\end{aligned}
\right.\label{8.299}
\end{equation}
To satisfy (\ref{8.299}) it is necessary to assume that the
coefficients $\alpha_0$ and $\alpha_4$ as in (\ref{8.294}) possess
the properties
\begin{equation}
\alpha_0(H)\rightarrow +\infty ,\ \ \ \ \alpha_4(T,H)\rightarrow
+\infty ,\ \ \ \ \text{as}\ h\rightarrow\infty .\label{8.300}
\end{equation}
Both conditions (\ref{8.294}) and (\ref{8.300}) are physical.

\subsection{Dynamic transitions}

In this subsection, to illustrate the main ideas, we only consider the case where $H$ is a
constant on $\Omega$. Therefore we shall study phase transition
dynamics of the ferromagnetic systems by using equations
(\ref{8.298}) for $h>0$. Analysis for more general case can be carried out in the same fashion, and will be reported elsewhere.

Take the transformation in (\ref{8.298})
\begin{equation}
M=M^*+M^{\prime}.\label{8.301}
\end{equation}
Then equations (\ref{8.298}) are rewritten as (drop the primes)
\begin{equation}
\left.
\begin{aligned}
&\frac{dM_1}{dt}=\beta_1M_1-2a_2M^2_1-a_2|M|^2-\alpha_4|M|^2M_1,\\
&\frac{dM_2}{dt}=\beta_2M_2-2a_2M_1M_2-\alpha_4|M|^2M_2,\\
&\frac{dM_3}{dt}=\beta_2M_3-2a_2M_1M_3-\alpha_4|M|^2M_3,
\end{aligned}
\right.\label{8.302}
\end{equation}
where
\begin{align*}
&  a_2=\alpha_4m_0-\beta h,\\
& \beta_1=\lambda_1+6\beta hm_0-3\alpha_4m^2_0-\alpha_2,\\
& \beta_2=2\beta hm_0-\alpha_4m^2_0-\alpha_2.
\end{align*}
The critical parameter curves $\beta_1=0$ and
$\beta_2=0$ are given by
\begin{align*}
& \beta_1=0\Rightarrow
T_1=T_0(H)+\frac{1}{\alpha_0}(\lambda_1+6\beta
hm_0-3\alpha_4m^2_0), \\
& \beta_2=0\Rightarrow T_2=T_0(H)+\frac{1}{\alpha_0}(2\beta
hm_0-\alpha_4m^2_0).
\end{align*}
 It is clear that $T_1>T_2$ provided
\begin{equation}
\lambda_1>2m_0(\alpha_4m_0-2\beta h),\ \ \ \ \text{for}\
h>0.\label{8.303}
\end{equation}

Therefore, under condition (\ref{8.303}),  the equations
(\ref{8.302}) have a transition at $T=T_1$ in the space
\begin{equation}
E=\{(M_1,0,0)|\ -\infty <M_1<\infty\}.\label{8.304}
\end{equation}
More precisely, we have the following transition theorem.

\bt\la{t8.21}
 Assume the condition (\ref{8.303}) and
$a_2\neq 0$. Then (\ref{8.302}) has a Type-III (mixed) transition at $T=T_1$, 
and the transition occurs in the space $E$. The phase
diagram is as shown in Figure \ref{f8.54}. Moreover we have the following
assertions:

\begin{itemize}
\item[(1)] There are two stable equilibrium states near $T=T_1$,
which are given by
\begin{align*}
& 
M^+_1
   =\left\{\begin{aligned} &  0 &&\text{if } T>T_1,\\
      & \frac{1}{2\alpha_4}[-3a_2+\sqrt{9a^2_2+4\alpha_4\beta_1}]
          && \text{if }   T<T_1,
\end{aligned}
\right.\\
&M^-_1=-\frac{1}{2\alpha_4}[3a_2+\sqrt{9a^2_2+4\alpha_4\beta_1}].
\end{align*}

\item[(2)] If  $T<T_1$, $M^+_1$ is stable in the  region $0<M_1<\infty$, and 
$M^-_1$ is stable in $-\infty <M_1<0$.

\item[(3)] If  $T>T_1$, $M^+_1=0$ is stable
in $-b<M_1<\infty$  and    $ M^-_1$ is stable in $-\infty <M_1<-b$, where
$$b=\frac{1}{2\alpha_4}[3a_2-\sqrt{9a^2_2+4\alpha_4\beta_1}]>0 \qquad \text{ for }T>T_1.$$
\end{itemize}
\et
\begin{figure}[hbt]
  \centering
  \includegraphics[width=.5\textwidth]{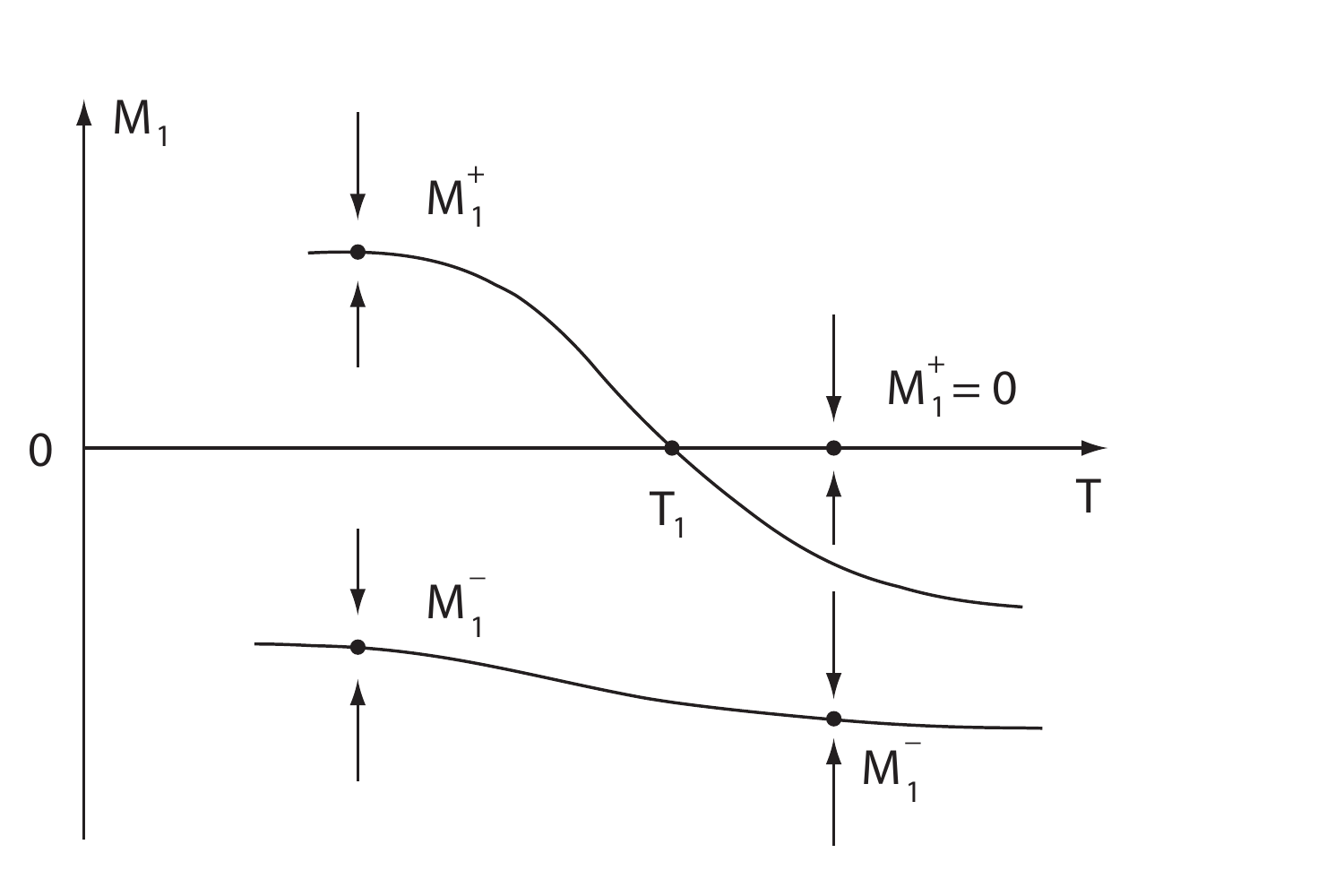}
  \caption{}\la{f8.54}
 \end{figure}

\bp
It is clear that if (\ref{8.303}) holds, then
\begin{align*}
&
\beta_1(T)\left\{\begin{aligned} 
& <0&& \text{ if } T>T_1,\\
& =0 && \text{ if }T=T_1,\\
& >0 && \text{ if }T<T_1,
\end{aligned}
\right.\\
&\beta_2(T_1)<0.
\end{align*}
Hence, by Theorem \ref{t5.1} the system (\ref{8.302}) has a transition at
$T=T_1$. Obviously, the space $E$ defined by (\ref{8.304}) is the
center manifold of (\ref{8.302}) near $T=T_1$. Hence, the reduced
equation  of (\ref{8.302}) on $E$ is expressed as
\begin{equation}
\frac{dM_1}{dt}=\beta_1M_1-3a_2M^2_1-\alpha_4M^3_1.\label{8.305}
\end{equation}
As $a_2\neq 0$, by Theorem \ref{t5.9} we infer from (\ref{8.305}) that
this transition is of type-III, and the transition solutions
satisfy
$$\alpha_4M^2_1+3a_2M_1-\beta_1=0.$$
By a direct compute one obtains Assertions (1) and (2). 

The proof is complete.
\ep

\section{Physical Conclusions  and Remarks}

\subsection{Physical predications based on Theorem \ref{t8.21}}

By (\ref{8.301}), the stable steady states of (\ref{8.298}) near
$T=T_1$ are
$$M^+=(m_0+M^+_1,0,0),$$
$$M^-=(m_0+M^-_1,0,0),$$
From  the  physical point of view, it should be
\begin{equation}
M^+_1\geq 0,\qquad  M^-_1<0, \qquad  m_0+M^-_1\geq 0.\label{8.306}
\end{equation}
The condition (\ref{8.306}) requires that
$$0<3a_2<\alpha_4m_0,$$
which is equivalent to
\begin{equation}
\beta h<\alpha_4m_0<\frac{3}{2}\beta h,\ \ \ \ (h>0)\label{8.307}
\end{equation}
where $m_0>0$ is a solution of the equation
\begin{equation}
\alpha_4m^3_0-3\beta hm^2_0+(\alpha_2-\lambda_1)m_0-h=0,\ \ \ \
(h>0),\label{8.308}
\end{equation}
near $T=T_1$.

Thus, the stable steady states $M^+$ and $M^-$ of (\ref{8.298})
near $T=T_1$ are physical provided that the cvoeficients
$\alpha_2(T,h),\alpha_4(T,h),\beta (T,h)$ and $\lambda_1(T,h)$
satisfy (\ref{8.307}) and (\ref{8.308}). In this case the
temperature $T_1$ is greater than the Curie temperature $T_c$:
$$T_1(H)>T_c=T_1(0)\qquad  \text{for}\ H\neq 0.$$

The two states $M^+$ and $M^-$ are mathematically equal, therefore
only by Theorem \ref{t8.21} we can not determine the magnetization
behaviors of ferromagnetic systems near $T=T_1$. However, we see
that the magnetization $M^+$ is stronger than $M^-$. Physically,
it implies that $M^+$ is favorable in $T<T_1$, and $M^-$ is in
$T>T_1$. Thus, from Theorem \ref{t8.21} there are two possible
magnetization behaviors, i.e., two magnetization functions:
\begin{align*}
&
\mu_1(T)=M^+(T)=(m_0(T)+M^+_1(T),0,0),\\
&
\mu_2(T)=\left\{\begin{aligned}
& M^+(T)=(m_0(T)+M^+_1(T),0,0)  && \text{ if } T<T_1,\\
& M^-(T)=(m_0(T)+M^-_1(T),0,0) && \text{ if } T\geq T_1.
\end{aligned}
\right.
\end{align*}
The function $\mu_1(T)$ is continuous on $T$, as shown in Figure
\ref{f8.55}(a), and its derivative is discontinuous at $T=T_1$:
$$\mu^{\prime}_1(T_1^-)-\mu^{\prime}_1(T_1^+)\simeq\left(\frac{1}{3}\frac{d}{dT}\frac{\beta_1(T_1)}
{\alpha_2(T_1)},0,0\right).
$$ 
The function $\mu_2(T)$  has a jump at
$T=T_1$, as shown in Figure \ref{f8.55}(b).
\begin{figure}[hbt]
  \centering
  \includegraphics[width=0.4\textwidth]{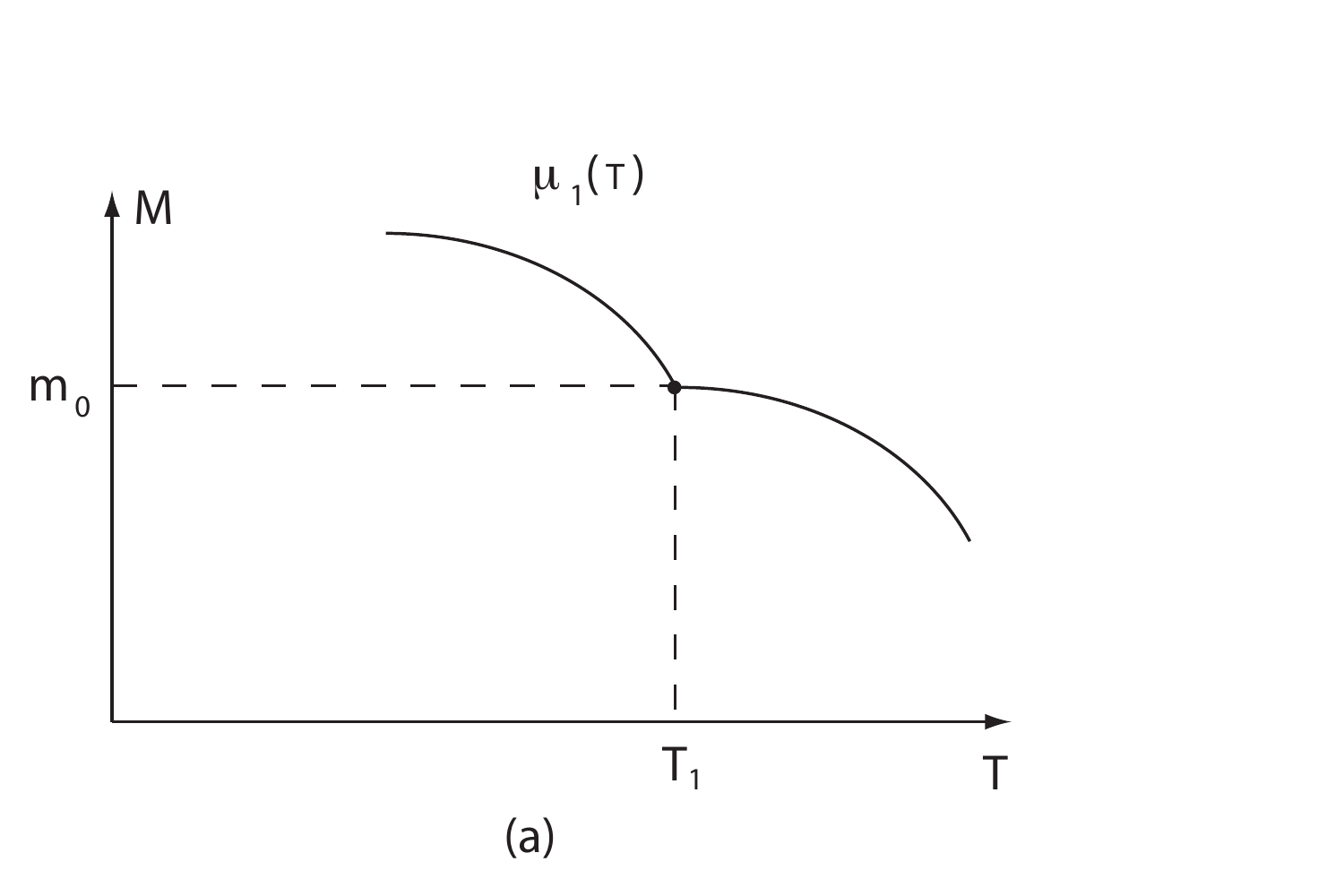}
  \includegraphics[width=0.4\textwidth]{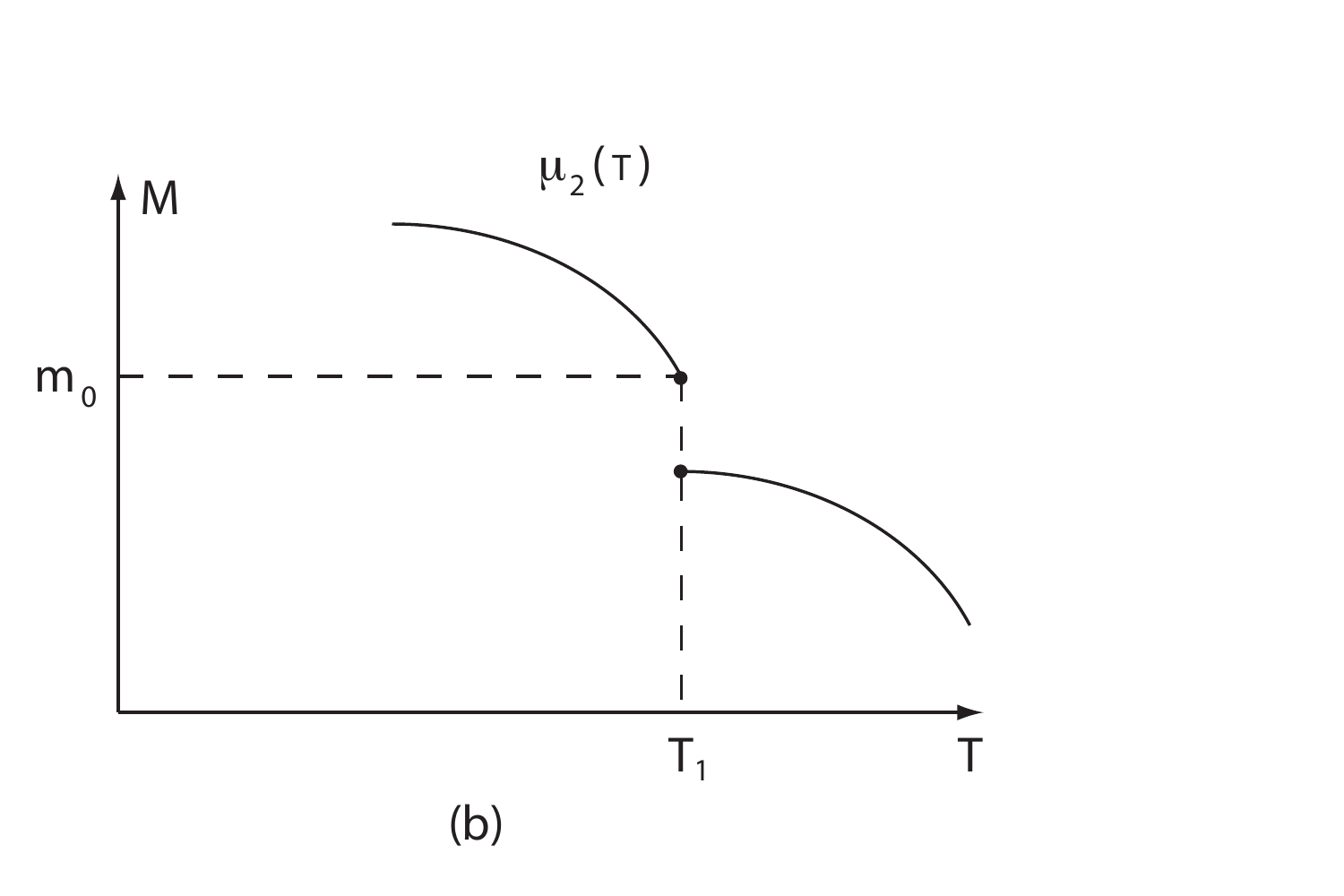}
  \caption{(a) The graph of function $\mu_1(T)$; (b) the
graph of function $\mu_2(T)$.}\la{f8.55}
 \end{figure}

On the other hand, by direct computation, the free energies of
$M^+$ and $M^-$ are shown to be given by  
\begin{align*}
&
G(M^+)=\left\{\begin{aligned} 
  &  G(m_0)  && \text{ if }  T\geq T_1,\\
& G(m_0)+\frac{1}{4}(M^+_1)^2(a_2M^+_1-\beta_1)  && \text{ if } T<T_1,
\end{aligned}
\right.\\
&
G(M^-)=G(m_0)+\frac{1}{4}(M^-_1)^2(a_2M^-_1-\beta_1),
\end{align*}
where $m_0$ is the magnetization induced by $H=(h,0,0)$ satisfying
(\ref{8.308}), and $\beta_1(T_1)=0$. It is clear that
\begin{equation}
G(M^+)>G(M^-)\ \ \ \ \text{near}\ T=T_1.\label{8.309}
\end{equation}

Hence, it follows from (\ref{8.309}) that the magnetization
behavior described by $\mu_2(T)$  is prohibited in real world
because the free energy can not abruptly increase (or decrease) in
a temperature decreasing (or increasing) process. Thus, by Theorem
\ref{t8.21} and (\ref{8.309}) we can derive the following physical
conclusion:

\bigskip

\noindent{\bf Physical Conclusion 5.1.}%8.8
{\it  When an external field $H$ is
present, the magnetization $M_H(T)$ of an isotropic ferromagnetic
system is continuous on the temperature $T$, and there is a
$T_1(H)>T_c$  ($T_c$ the Curie temperature) with $T_1(H)\rightarrow
T_c$ as $H\rightarrow 0$ such that $M_H(T)$ is not differentiable
at $T=T_1$, whose derivative has a finite jump
$$
M^{\prime}_H(T_1^-)-M^{\prime}_H(T_1^+)=a>0\ \ \ \ (a<\infty).
$$ 
Moreover, the graph of $M_H(T)=\mu_1(T)$ as shown in Figure
\ref{f8.55}(a), and $M_H(T)\rightarrow M_0(T)$ as $H\rightarrow 0$ with
$$
M_0(T)=\left\{\begin{aligned}
& 0    && \text{ if } T\geq T_c,\\
& M_s(T)  && \text{ if }T<T_c,
\end{aligned}
\right.
$$ 
where $M_s(T)$ is the spontaneous magnetization (see
Figure \ref{f8.52}).
}

\subsection{Asymmetry of fluctuations}
The above discussions suggest that for ferromagnetic systems, there are two
possible phase transition behaviors near a critical point, and
theoretically each of them has some probability to take place,
however only one of them can appear in reality. For the
ferromagnetic systems we again see this situation. This phenomena is also observed in  phase transitions for PVT systems \cite{mw-pvt}.

One explanation of such  phenomena is that  the symmety of fluctuation near
a critical point is not generally true in equilibrium phase
transitions. To make the statement more clear, we first introduce
some concepts.

Let $G(u,\lambda )$ be free energy of a thermodynamic system,
$u=(u_1,\cdots,u_n)$ be the order parameter, and $\lambda
=(\lambda_1,\cdots,\lambda_m)$ the control parameter $(n,m\geq
1)$. Assume that $u$ is defined in the function space $L^2(\Omega
,\R^n)$ and $\lambda\in \R^m$. Then the space
$$X=\{(u,\lambda )|\ u\in L^2(\Omega ,\R^n), \lambda\in \R^m\}$$
is called the state space of the system.

Let $(u_0,\lambda_0)\in X$ be a stable equilibrium state of the
system; namely $(u_0,\lambda_0)$ is a locally minimal state of
$G(u,\lambda )$. We say that the system has a fluctuation at
$(u_0,\lambda_0)$ if it deviates randomly from $(u_0,\lambda_0)$
to $(\widetilde{u},\widetilde{\lambda})$ with
$$\|\widetilde{u}-u_0\|+|\widetilde{\lambda}-\lambda_0|>0.$$
In this case, $(\widetilde{u},\widetilde{\lambda})$ is called a
state of fluctuation.

The so called symmetry of fluctuation means that for given $r>0$,
all states $(\widetilde{u},\widetilde{\lambda})$ of fluctuation
satisfying
$$
\|\widetilde{u}-u_0\|+|\widetilde{\lambda}-\lambda_0|=r,\ \ \ \
(\widetilde{u},\widetilde{\lambda})\in X,
$$ 
have the same probability to appear in real world. Otherwise, we say that the
fluctuation is asymmetric.

The observations in both the PVT systems and 
the ferromagnetic systems strongly suggest the following  physical conjecture, regarding to  the uniqueness of
transition behaviors.

\bigskip
\noindent
{\bf  Physical Conjecture} (Asymmetry of Fluctuations). 
{\it 
 The symmetry of fluctuations for general thermodynamic systems may not
be universally true. In other words, in some systems with
multi-equilibrium states, the fluctuations near a critical point
occur only in one basin of attraction of some equilibrium states,
which are the ones that can be physically observed.}

\appendix

\section{Recapitulation of the Dynamic Transition Theory}
In this appendix we recall some basic elements of the dynamic transition theory developed by the authors \cite{b-book, chinese-book}, which are used to carry out the dynamic transition analysis for the ferromagnetism systems in this article. 

Let $X$  and $ X_1$ be two Banach spaces, $X_1\subset X$ a compact and
dense inclusion. In this chapter, we always consider the following
nonlinear evolution equations
\begin{equation}
\left. 
\begin{aligned} 
&\frac{du}{dt}=L_{\lambda}u+G(u,\lambda),\\
&u(0)=\varphi ,
\end{aligned}
\right.\label{5.1}
\end{equation}
where $u:[0,\infty )\rightarrow X$ is unknown function,  and 
$\lambda\in \R^1$  is the system parameter.

Assume that $L_{\lambda}:X_1\rightarrow X$ is a parameterized
linear completely continuous field depending contiguously on
$\lambda\in \R^1$, which satisfies
\begin{equation}
\left. 
\begin{aligned} 
&L_{\lambda}=-A+B_{\lambda}   && \text{a sectorial operator},\\
&A:X_1\rightarrow X   && \text{a linear homeomorphism},\\
&B_{\lambda}:X_1\rightarrow X&&  \text{a linear compact  operator}.
\end{aligned}
\right.\label{5.2}
\end{equation}
In this case, we can define the fractional order spaces
$X_{\sigma}$ for $\sigma\in \R^1$. Then we also assume that
$G(\cdot ,\lambda ):X_{\alpha}\rightarrow X$ is $C^r(r\geq 1)$
bounded mapping for some $0\leq\alpha <1$, depending continuously
on $\lambda\in \R^1$, and
\begin{equation}
G(u,\lambda )=o(\|u\|_{X_{\alpha}}),\ \ \ \ \forall\lambda\in
\R^1.\label{5.3}
\end{equation}

Hereafter we always assume the conditions (\ref{5.2}) and
(\ref{5.3}), which represent that the system (\ref{5.1}) has
a dissipative structure.

In the following we introduce the definition of transitions for
(\ref{5.1}).

\bd\la{d5.1}
We say that the system (\ref{5.1}) has a
transition of equilibrium from $(u,\lambda )=(0,\lambda_0)$ on
$\lambda >\lambda_0$ (or $\lambda <\lambda_0)$ if  the following two conditions are 
satisfied:

\begin{itemize}

\item[(1)] when $\lambda <\lambda_0$ (or $\lambda >\lambda_0),
u=0$ is locally asymptotically stable for (\ref{5.1}); and

\item[(2)] when $\lambda >\lambda_0$ (or $\lambda <\lambda_0)$,
there exists a neighborhood $U\subset X$ of $u=0$ independent of
$\lambda$, such that for any $\varphi\in U \setminus \Gamma_{\lambda}$ the
solution $u_{\lambda}(t,\varphi )$ of (\ref{5.1}) satisfies that
$$\left. 
\begin{aligned}
&\limsup_{t\rightarrow\infty}\|u_{\lambda}(t,\varphi
)\|_X\geq\delta (\lambda )>0,\\
&\lim_{\lambda\rightarrow\lambda_0}\delta(\lambda )\geq 0,
\end{aligned}
\right.$$ 
where $\Gamma_{\lambda}$ is the stable manifold of
$u=0$, with  codim $\Gamma_{\lambda}\geq 1$ in $X$
for $\lambda >\lambda_0$ (or $\lambda <\lambda_0)$.
\end{itemize}
\ed

Obviously, the attractor bifurcation of (\ref{5.1}) is a type of
transition. However,  bifurcation and
transition are two different, but related concepts. 
Definition \ref{d5.1} defines the transition of (\ref{5.1}) from a stable
equilibrium point to other states (not necessary equilibrium state).
In general, we can define transitions from one attractor to
another as follows.

\bd\la{d 5.2}
Let $\Sigma_{\lambda}\subset X$ be an
invariant set of (\ref{5.1}). We say that (\ref{5.1}) has a
transition of states from $(\Sigma_{\lambda_0},\lambda_0)$ on
$\lambda >\lambda_0$ (or $\lambda <\lambda_0)$ if the following conditions are 
satisfied:

\begin{itemize}

\item[(1)] when $\lambda <\lambda_0$ (or $\lambda >\lambda_0),
\Sigma_{\lambda}$ is a local minimal attractor, and 

\item[(2)]
when $\lambda >\lambda_0$ (or $\lambda <\lambda_0)$, there exists
a neighborhood $U\subset X$ of $\Sigma_{\lambda}$ independent of
$\lambda$ such that for any $\varphi\in
U \setminus (\Gamma_{\lambda}\cup\Sigma_{\lambda})$, 
the solution $u(t,\varphi)$ of (\ref{5.1}) satisfies that
$$\left. 
\begin{aligned}
&\limsup_{t\rightarrow\infty}\text{\rm dist}(u(t,\varphi),\Sigma_{\lambda})
    \geq\delta (\lambda )>0,\\
&\lim\limits_{\lambda\rightarrow\lambda_0}\delta (\lambda
)=\delta\geq 0,
\end{aligned}
\right.$$ 
where $\Gamma_{\lambda}$ is the stable manifolds of
$\Sigma_{\lambda}$ with codim $\Gamma_{\lambda}\geq 1$.
\end{itemize}
\ed

 Let the eigenvalues (counting multiplicity) of $L_{\lambda}$ be given by
$$\{\beta_j(\lambda )\in \C\ \   |\ \ j=1,2,\cdots\}$$
Assume that
\begin{align}
&  \text{Re}\ \beta_i(\lambda )
\left\{ 
 \begin{aligned} 
 &  <0 &&    \text{ if } \lambda  <\lambda_0,\\
& =0 &&      \text{ if } \lambda =\lambda_0,\\
& >0&&     \text{ if } \lambda >\lambda_0,
\end{aligned}
\right.   &&  \forall 1\leq i\leq m,  \label{5.4}\\
&\text{Re}\ \beta_j(\lambda_0)<0 &&  \forall j\geq
m+1.\label{5.5}
\end{align}

The following theorem is a basic principle of transitions from
equilibrium states, which provides sufficient conditions and a basic
classification for transitions of nonlinear dissipative systems.
This theorem is a direct consequence of the center manifold
theorems and the stable manifold theorems; we omit the proof.

\bt\la{t5.1}
 Let the conditions (\ref{5.4}) and
(\ref{5.5}) hold true. Then, the system (\ref{5.1}) must have a
transition from $(u,\lambda )=(0,\lambda_0)$, and there is a
neighborhood $U\subset X$ of $u=0$ such that the transition is one
of the following three types:

\begin{itemize}
\item[(1)] {\sc Continuous Transition}: 
there exists an open and dense set
$\widetilde{U}_{\lambda}\subset U$ such that for any
$\varphi\in\widetilde{U}_{\lambda}$,  the solution
$u_{\lambda}(t,\varphi )$ of (\ref{5.1}) satisfies
$$\lim\limits_{\lambda\rightarrow\lambda_0}\limsup_{t\rightarrow\infty}\|u_{\lambda}(t,\varphi
)\|_X=0.$$ In particular, the attractor bifurcation of (\ref{5.1})
at $(0,\lambda_0)$ is a continuous transition.

\item[(2)] {\sc Jump Transition}: 
for any $\lambda_0<\lambda <\lambda_0+\varepsilon$ with some $\varepsilon >0$, there is an open
and dense set $U_{\lambda}\subset U$ such that 
for any $\varphi\in U_{\lambda}$, 
$$\limsup_{t\rightarrow\infty}\|u_{\lambda}(t,\varphi
)\|_X\geq\delta >0,$$ 
where $\delta >0$ is independent of $\lambda$. 
This type of transition  is also called the discontinuous 
transition. 

\item[(3)] {\sc Mixed Transition}: 
for any $\lambda_0<\lambda <\lambda_0+\varepsilon$  with some $\varepsilon >0$, 
$U$ can be decomposed into two open sets
$U^{\lambda}_1$ and $U^{\lambda}_2$  ($U^{\lambda}_i$ not necessarily
connected):
$$\bar{U}=\bar{U}^{\lambda}_1+\bar{U}^{\lambda}_2,\ \ \
\ U^{\lambda}_1\cap U^{\lambda}_2=\emptyset ,$$ 
such that
\begin{align*}
&\lim\limits_{\lambda\rightarrow\lambda_0}\limsup_{t\rightarrow\infty}\|u(t,\varphi
)\|_X=0   &&   \forall\varphi\in U^{\lambda}_1,\\
& \limsup_{t\rightarrow\infty}\|u(t,\varphi
)\|_X\geq\delta >0 && \forall\varphi\in U^{\lambda}_2.
\end{align*}
%where  $U^{\lambda}_1$ is called the stable domain,  and  $U^{\lambda}_2$
%is the unstable domain.
\end{itemize}
\et

An important aspect of the  transition theory is to determine which 
of the three types of transitions given by Theorem \ref{t5.1} occurs in
a specific  problem. We refer the interested readers to \cite{chinese-book, b-book} for more discussions. 
Instead, here we consider the transition of (\ref{5.1}) from a simple critical
eigenvalue. Let the eigenvalues $\beta_j(\lambda )$ of
$L_{\lambda}$ satisfy
\begin{equation}
\left. 
\begin{aligned} 
&  
\beta_1(\lambda )
\left\{
 \begin{aligned}
 & <0 &&\text{ if } \lambda <\lambda_0,\\
& =0 &&\text{ if } \lambda =\lambda_0,\\
& >0 &&\text{ if } \lambda >\lambda_0,
\end{aligned}
\right.\\
& \text{Re}\beta_j(\lambda_0)<0 &&\forall j\geq 2,
\end{aligned}
\right.\label{5.35}
\end{equation}
where $\beta_1(\lambda )$ is a real eigenvalue.

Let $e_1(\lambda )$ and $e^*_1(\lambda )$   be  the eigenvectors of
$L_{\lambda}$ and $L^*_{\lambda}$ respectively corresponding to
$\beta_1(\lambda )$ with
$$L_{\lambda_0}e_1=0,\ \ \ \ L^*_{\lambda_0}e^*_1=0,\ \ \ \
<e_1,e^*_1>=1.$$ 
Let $\Phi (x,\lambda )$    be the center manifold
function of (\ref{5.1}) near $\lambda =\lambda_0$. We assume that
\begin{equation}
<G(xe_1+\Phi (x,\lambda_0),\lambda_0),e^*_1>=\alpha
x^k+o(|x|^k),\label{5.36}
\end{equation}
where $k\geq 2$ an integer and $\alpha\neq 0$ a real number.

We have the following transition theorems.
 \begin{figure}%[hbtp]
  \centering
  \includegraphics[width=0.32\textwidth]{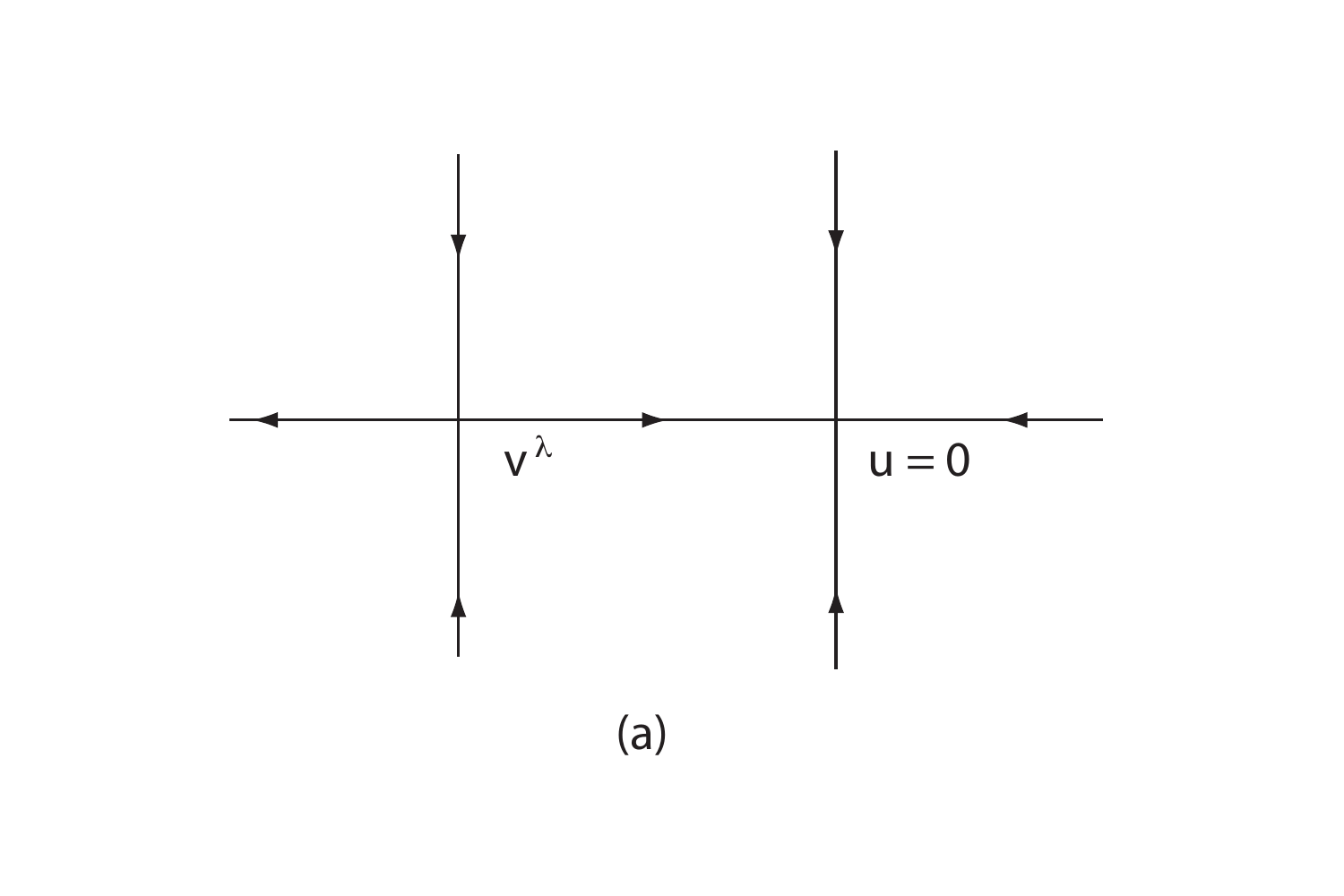}
   \includegraphics[width=0.22\textwidth]{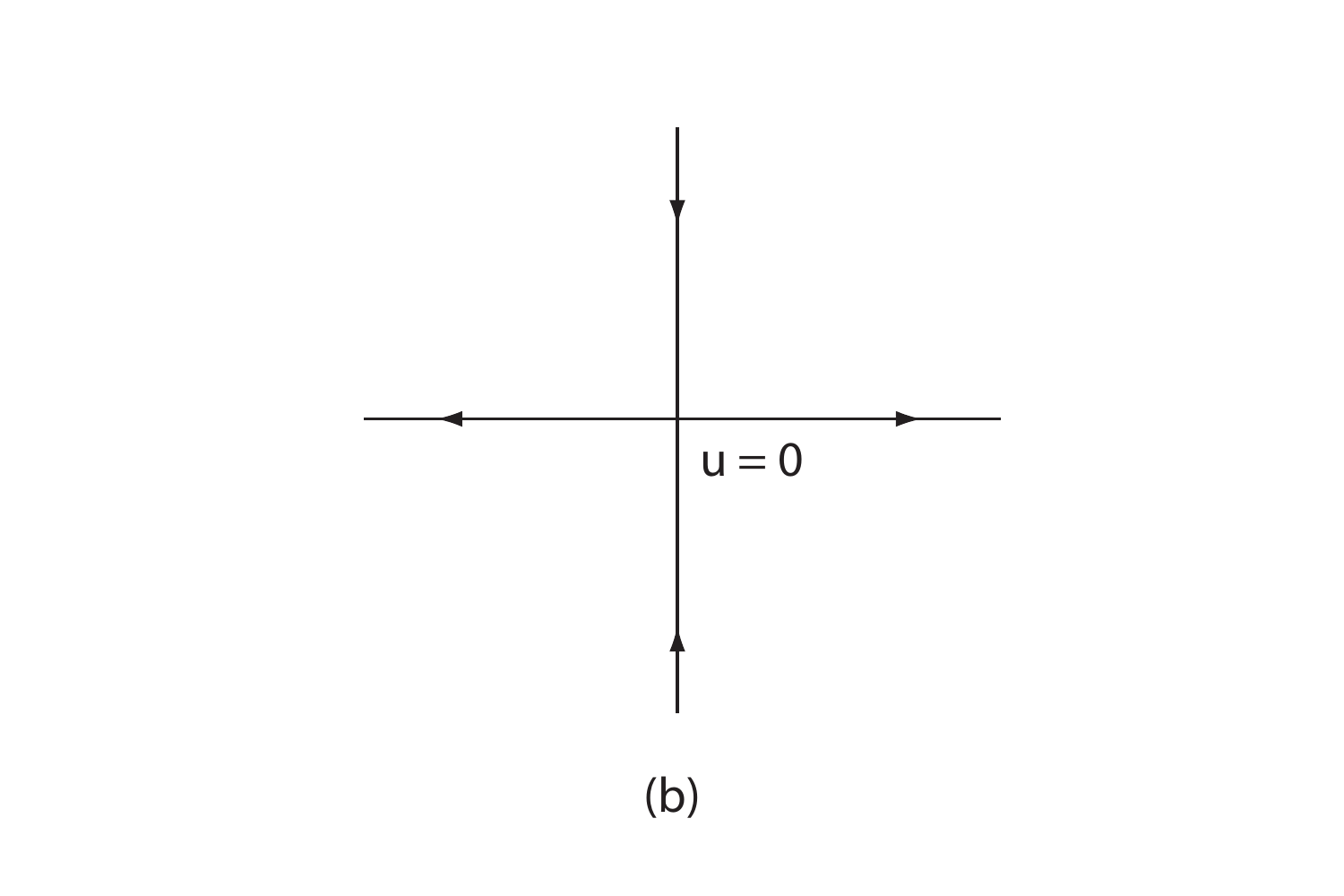} 
   \includegraphics[width=0.32\textwidth]{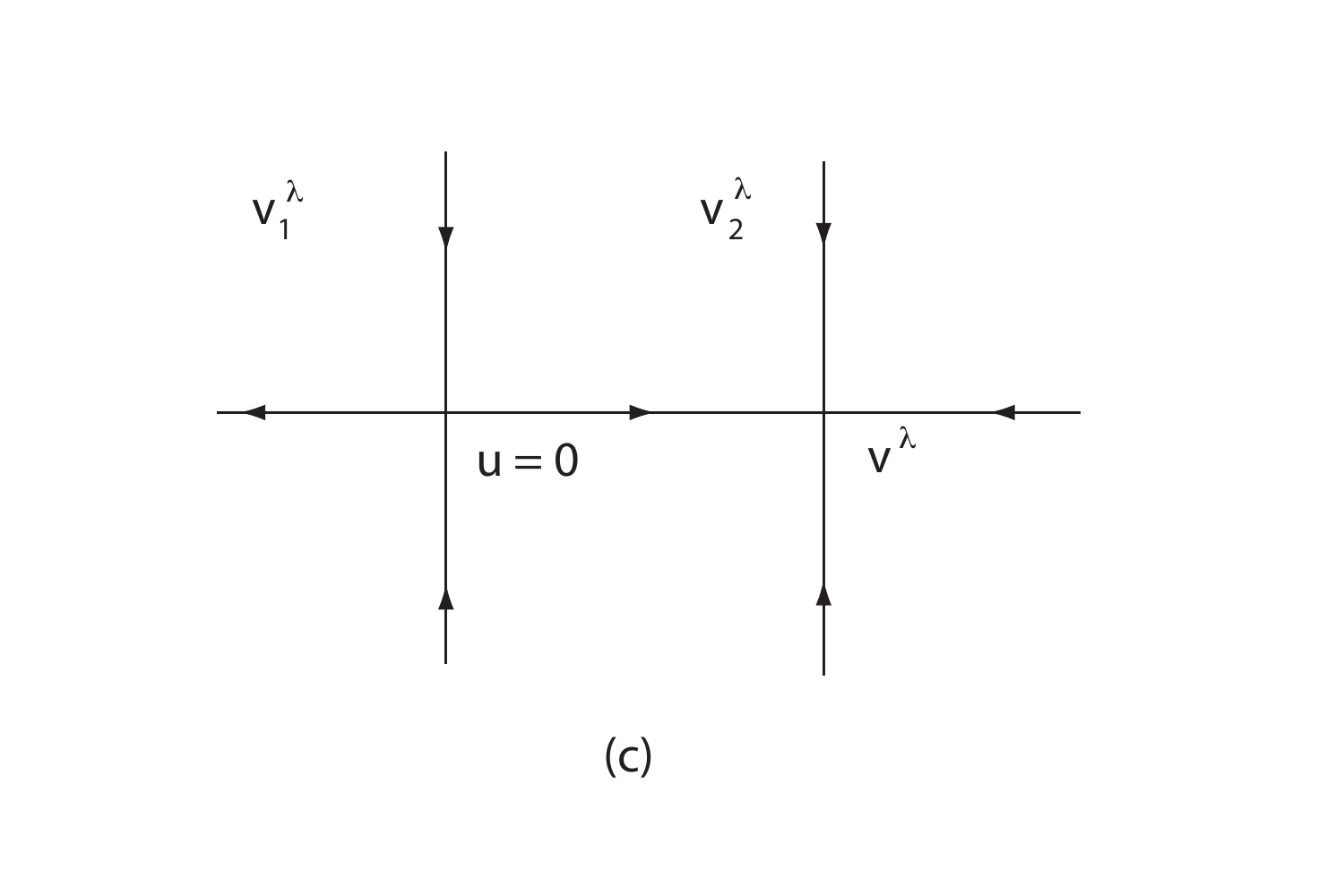}
  \caption{Topological structure of the mixing transition of
(\ref{5.1}) when $k$=even and $\alpha\neq 0$: (a) $\lambda
<\lambda_0$; (b) $\lambda =\lambda_0$; (c) $\lambda >\lambda_0$. Here
$U^{\lambda}_1$ is the unstable domain, and $U^{\lambda}_2$ the
stable domain.}\la{f5.7}
 \end{figure}

\bt\la{t5.9}
 Under the conditions (\ref{5.35}) and
(\ref{5.36}), if $k$=even and $\alpha\neq 0$, then we have the
following assertions:

\begin{enumerate}

\item (\ref{5.1}) has a mixed transition from
$(0,\lambda_0)$. More precisely, there exists a neighborhood
$U\subset X$ of $u=0$ such that $U$ is separated into two disjoint
open sets $U^{\lambda}_1$ and $U^{\lambda}_2$ by the stable
manifold $\Gamma_{\lambda}$ of $u=0$ satisfying the following properties:

\begin{enumerate}

\item $U=U^{\lambda}_1+U^{\lambda}_2+\Gamma_{\lambda}$,

\item the transition in $U^{\lambda}_1$ is jump, and 

\item the transition in $U^{\lambda}_2$ is
continuous. The local transition structure is as shown in Figure \ref{f5.7}.

\end{enumerate}

\item (\ref{5.1}) bifurcates in $U^{\lambda}_2$ to a unique
singular point $v^{\lambda}$ on $\lambda >\lambda_0$, which is an
attractor such that for any $\varphi\in U^{\lambda}_2$, 
$$\lim\limits_{t\rightarrow\infty}\|u(t,\varphi
)-v^{\lambda}\|_X=0,$$
where $u(t,\varphi )$ is the solution of (\ref{5.1}). 

\item (\ref{5.1})\ bifurcates on $\lambda <\lambda_0$ to a unique saddle
point $v^{\lambda}$ with the Morse index one. 

\item The bifurcated singular point $v^{\lambda}$ can be expressed as
$$v^{\lambda}=-(\beta_1(\lambda )/\alpha
)^{{1}/{(k-1)}}e_1+o(|\beta_1|^{{1}/{(k-1)}}).$$
\end{enumerate}
\et

\bibliographystyle{siam}

\bibliography{ferromagnetism}

\def\cprime{$'$}
\begin{thebibliography}{1}

\bibitem{b-book}
{\sc T.~Ma and S.~Wang}, {\em Bifurcation theory and applications}, vol.~53 of
  World Scientific Series on Nonlinear Science. Series A: Monographs and
  Treatises, World Scientific Publishing Co. Pte. Ltd., Hackensack, NJ, 2005.

\bibitem{mw-pvt}
\leavevmode\vrule height 2pt depth -1.6pt width 23pt, {\em Dynamic phase
  transitions in pvt systems}, submitted,  (2007).

\bibitem{chinese-book}
\leavevmode\vrule height 2pt depth -1.6pt width 23pt, {\em Stability and
  Bifurcation of Nonlinear Evolution Equations}, Science Press, 2007.

\bibitem{onuki}
{\sc A.~Onuki}, {\em Phase transition dynamics}, Cambridge University Press,
  2007.

\bibitem{reichl}
{\sc L.~E. Reichl}, {\em A modern course in statistical physics}, A
  Wiley-Interscience Publication, John Wiley \& Sons Inc., New York,
  second~ed., 1998.

\end{thebibliography}
\end{document}